\documentclass[conference]{IEEEtran}
\usepackage{cite}
\usepackage{graphicx}
\usepackage{comment}
\usepackage{balance}
\usepackage{acronym}
\usepackage{caption}
\usepackage{subcaption}
\usepackage{amsmath}
\usepackage{tabularx}
\usepackage{multirow}
\usepackage{array}
\usepackage{booktabs}

\acrodef{CNN}{Convolutional Neural Network}
\acrodef{VAD}{Voice Activity Detection}
\acrodef{SNR}{Signal to Noise Ratio}
\acrodef{LRT}{Likelihood Ratio Test}
\acrodef{DNN}{Deep Neural Network}
\acrodef{ROC}{Receiver Operating Characteristic}
\acrodef{RNN}{Recurrent Neural Network}
\acrodef{RIR}{Room Impulse Response}
\acrodef{FC}{Fully Connected}
\acrodef{SGD}{Stochastic Gradient Descent}
\acrodef{BCE}{Binary Cross Entropy}
\acrodef{AUC}{Area Under the Curve}
\acrodef{EER}{Equal Error Rate}
\acrodef{FCNN}{Fully Convelotional Neural Network}
\acrodef{ACAM}{Adaptive Context Attention Model}
\acrodef{SA}{Self Attention}

\acrodef{LSTM}{Long Short-Term Memory}
\acrodef{bDNN}{boosted DNN}
\acrodef{ASR}{Automatic Speech Recognition}
\acrodef{SOTA}{State Of The Art}
\acrodef{ATF}{Acoustic Transfer Function}
\acrodef{NLP}{Natural Language Processing}

\ifCLASSINFOpdf

\else

\fi

\begin{document}

\newcommand{\noteb}[1]{\textcolor{blue}{#1}}
\newcommand{\noter}[1]{\textcolor{red}{#1}}
\newcommand{\noteg}[1]{\textcolor{green}{#1}}
\newcommand{\noteo}[1]{\textcolor{olive}{#1}}
\newcommand{\notem}[1]{\textcolor{magenta}{#1}}

\def\y{y(t,k)}
\def\sd{s_d(t,k)}
\def\rd{r_d(t,k)}
\def\si{s_i(t,k)}
\acrodef{STFT}{short-time Fourier transform}
\acrodef{ISTFT}{inverse short-time Fourier transform}
\acrodef{BSS}{Blind Source Separation}
\acrodef{DOA}{Direction Of Arrival}

\title{CNN self-attention voice activity detector}

\author{\IEEEauthorblockN{Amit Sofer}
\IEEEauthorblockA{OriginAI}
\and

\and
\IEEEauthorblockN{ Shlomo E. Chazan}
\IEEEauthorblockA{OriginAI}}

\maketitle

\begin{abstract}
In this work we present a novel single-channel \ac{VAD} approach. We utilize a \ac{CNN} which exploits the  spatial  information  of  the noisy input spectrum to extract frame-wise embedding sequence, followed by a \ac{SA} Encoder with a goal of finding  contextual information from the embedding sequence. Different from  previous  works  which  were  employed  on  each  frame  (with context  frames)  separately,  our  method  is  capable  of  processing the  entire  signal  at  once, and thus enabling  long  receptive  field.
We show  that  the  fusion  of  CNN  and  SA  architectures  outperforms methods based solely on \ac{CNN} and \ac{SA}.
Extensive experimental-study shows that our model outperforms previous models on real-life benchmarks, and provides \ac{SOTA} results with relatively small and lightweight model.
\end{abstract}

\IEEEpeerreviewmaketitle

\section{Introduction}
Accurately detecting the presence or absence of speech in a noisy single-channel recording is still an open issue. The \ac{VAD} is commonly used as a trigger for different audio processing algorithms, such as speech communication systems, \ac{ASR} and echo cancellation. In a noisy and/or  reverberant enclosure, the challenge is even more difficult.

Classical signal-processing-based approaches use acoustic features such
as zero-crossing rate, pitch detection, and energy thresholds to determine the presence of speech~\cite{benyassine1997itu, renevey2001entropy, dhananjaya2009voiced}. Yet, relying on a pre-defined threshold, in real-life scenarios these methods performs poorly.
A statistical model was afterwards used to model the speech and noise signals in~\cite{ramirez2007voice, sohn1999statistical}. Their parameters were learned utilizing the \ac{LRT} to determine the most likely hypothesis.
Although, a noticeable improvement was gained,  these methods are heavily susceptible to non-stationary noise and reverberation conditions.

In recent years, \ac{DNN}-based detectors were introduced utilizing dense-networks, \acp{CNN} and \acp{RNN}~\cite{zhang2015boosting, ryant2013speech,thomas2014analyzing, zazo2016feature,shannon2017improved, gelly2017optimization}. These methods have shown promising results compared to the classical algorithms.  
\acp{CNN} are very powerful and widely used models, since they extract useful features from images (including spectrums), which were found beneficial in many downstream tasks including speech-processing algorithms~\cite{chazan2021speech, wang2018supervised}.
Typically, a CNN-based VAD system requires employing a flattening operation on the output of the stacked \ac{CNN} layers, followed by several \ac{FC} layers.  This approach has several drawbacks. First, each additional \ac{FC} layers adds a significant number of trainable parameters. Second, this approach requires a fixed size inputs which tends to be relatively small, resulting in applying the VAD continuously and separately on small parts of the inputs.  
In addition, the input length constraint prevents long receptive information which might be beneficial for this task.
A \ac{FCNN} architecture can be used to overcome these hurdles, however, the receptive field of this network depends on the depth of the \ac{CNN}, which is still limited.

One of the first attempts for an attention-based \ac{VAD}, the \ac{ACAM}, was introduced in \cite{kim2018voice}. Their suggested approach consist of a \ac{LSTM} and an attention mechanism which allowed the model to focus on relevant input segments. However, Its sequential nature prevent different input segments from relating to each other, which may lead to information loss and catastrophic forgetting. Inspired by the prominence of Self-Attention architecture~\cite{vaswani2017attention} that is sweeping through most AI-related domains \cite{zhang2019self, lee2019self, dosovitskiy2020image},  a \ac{SA}-based encoder \ac{VAD} was presented in~\cite{jo2021self}.  The architecture of this model is constructed mainly on \ac{SA} blocks. The idea behind this architecture is to model the contextual information between the acoustic input-frames. Unfortunately, the \ac{SA} mechanism is used only on short context frames. Furthermore, without using a \ac{CNN} on the input,  the spectral information of the speech might not be fully exploited.

In this paper, we present a VAD which consist of the fusion of a \ac{CNN} embedder followed by a self-attention encoder.
The \ac{CNN} is built to preserve the frame information and extract frame-wise features, while the \ac{SA}-encoder process the entire embedded sequence in order to determine which of all input frames are relevant for classifying each of the frames. 
We show that the combined architectures performs better than each of its components separately. Furthermore, we show that the proposed methods not only achieve \ac{SOTA} results on various benchmarks, but also computationally efficient as it is process the entire signal in a single step.    


\section{Problem Formulation}
Let us assume that a single microphone input signal $x(t)$ is modeled as 
\begin{equation}
x(t) =  s(t)*h(t)+n(t)
\label{eq:prooblem_time}
\end{equation}
where $s(t)$ is a non-reverberant speech signal, $h(t)$ is the \ac{RIR} between the speaker and the microphone, $*$ is the convolution operator and $n(t)$ is an additive noise signal. 

In the \ac{STFT} domain, the convolution is approximated as a multiplication operation, and \eqref{eq:prooblem_time} can be rewritten as,
\begin{equation}
x(l,f) = I(l)\cdot s(l,f) \cdot h(l,f)+n(l,f)
\label{eq:prooblem_stft}
\end{equation}

\begin{figure*}[h]
    \centering
    \includegraphics[width=0.99\textwidth]{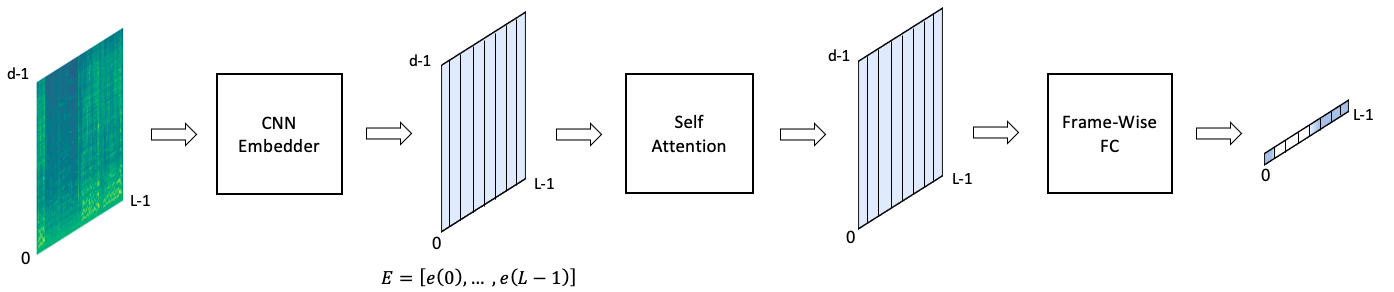}
    \caption{The Proposed Architecture}
    \label{fig:diagram}
\end{figure*}

where $l \in \{0,\ldots, L-1\}$ and $f \in \{0,\ldots, F-1\}$ are the frame and frequency indexes, respectively. The term $I(l)$ indicates the activity of $s(l,f)$ with,
\begin{equation}
    I(l)=
        \begin{cases}
                0 & \quad \sum_{f=0}^{F-1} s(l,f)<\text{Th} \\
                1 & \quad \sum_{f=0}^{F-1} s(l,f)\geq \text{Th}
        \end{cases}
    \label{eq:indicator}
\end{equation} 
where $\text{Th}$ is a pre-defined threshold applied on the clean signals.

The \ac{VAD} goal is to accurately detect frames in which the speaker is active given a noisy reverberant signal, 
\begin{equation}
    V(l)=p(I(l)=1|x).
    \label{eq:vad}
\end{equation}

\section{Proposed Model}
\label{sec:encoder}

A block diagram of the proposed model is depicted in Fig.~\ref{fig:diagram}. The Mel-spectrum of the noisy input, $x$, is used as an input to our model. It is first propagates through a CNN, while preserving the input length dimension, $L$. The CNN 'embedder' takes the local spatial information of the input spectrum into account, and outputs frame-wise embedding, $e(l)$. The embedding sequence, $E=[e(0),\ldots,e(L-1)]$, is then fed into a \ac{SA} encoder. The \ac{SA} is responsible for calculating the attention of all other embedding with respect to $e(l)$. Finally, a frame wise \ac{FC} layer is used to estimate \eqref{eq:vad}.



\subsection{Self-attention encoder}
\ac{SA} mechanism~\cite{vaswani2017attention} is at the core of prominent architectures in multiple Machine Learning domains such as \ac{NLP} and Computer Vision (CV)~\cite{dosovitskiy2020image, liu2021swin}. The main gits of the \ac{SA} encoder is mapping a query and a set of key-value pairs to an output. The output is computed as a weighted sum of the values, where the weight assigned to each value is computed using the given query with the corresponding key.
In our case, given the frame-wise embedding sequence $E \in \mathcal{R}^{d\times L}$ where $L$ is the input sequence length and $d$ is the dimension of each element of the sequence,
a single layer attention of the multi-head attention with $H$ attention heads is computed as 
\begin{equation}
\text{MultiHead} = Concat(head_0,\ldots,head_{H-1})W^O
\end{equation}
where 
\begin{equation}
\begin{aligned}
head_i & = softmax(\frac{Q_i K_i^T}{\sqrt{d}})V_i,
\\
Q_i  & = E\cdot W_i^{Q} , \hspace{5pt}  K_i = E\cdot W_i^{K}, \hspace{5pt} V = E\cdot W_i^{V}
\end{aligned}
\end{equation}
The terms $W_i^{Q}, W_i^{K}, W_i^{V}, W^{O} \quad i \in [0, \dots, H-1 ]$ are the learned parameter-sets of the multi-head attention layer.

The average attention ( across all $H$ heads) from each frame to each other frame can be written as,
\begin{equation}
\label{eq:AverageAttention}
\text{AverageAttention} = \frac{1}{H} \sum \limits_{i=0}^{H-1} softmax(\frac{Q_iK_i^T}{\sqrt{d}}).
\end{equation}

Note, that the average attention dimensions are $[L,L]$, connecting each frame information with all other frames.

After the self-attention layer, a residual connections and layer-norm are applied and then a feed-forward layer, which first increases the dimensions of each input vector from $d$ to $d_{ff}$, and then decreases it to the original size, followed by residual connections together with additional layer-norm.
\subsection{Model Architecture}
We construct a four layer \ac{CNN} with batch-norm, PReLU non-linearly and max-polling between each of the convolution layers.
The  \ac{CNN} output dimentions are $[L,F',C]$, where $F'$ represents the remaining frequency dimension (after the pooling operations) and $C$ represents the number of channels in the CNN output layer.

Frame-wise flattening operation is carried out to the output of the CNN. The remaining frequencies $F'$ and channels $C$, are flatten for each frame while preserving the frame dimension, meaning we get a $[L,F' \cdot C]$ embedding.
We then use a frame-wise \ac{FC} layer that transforms each embedding from size $F' \cdot C$ to size $d$, resulting in frame-wise embedding, $E \in \mathcal{R}^{d\times L}$.

The \ac{SA} encoder is then applied on the embedding sequence. The output of the \ac{SA} encoder is also a sequence, but unlike the input sequence, the output sequence contains contextual information from the entire input. The output sequence passes through a frame-wise \ac{FC} layer which finally detects the speech presence of each frame.
A moving average is used for further smoothing of the model outputs.

As mentioned above, the Mel-spectrum is used as the input to the network, with window length of 1024 bins, and hop size of 512. With sample-rate of $fs=8$~KHz, the frame and hop lengths are $128$ and $64$~ms, respectively. The number of Mel-filters was set to $F=256$ and varying length $L$, resulting in a input size of $L,F$.
In the \ac{CNN}, we use a $3 \times 3$ kernels with $C=32$ channels across all layers and we use a $[2,1]$ max polling operation between each convolution layer. After 4 convolution layers we get $F'=16$, meaning after the frame-wise flattening operation we get a feature vector of length $F' \cdot C=512$ for each frame, which is based not only of the single frame features, but also on features of proximate frames.
In the encoder layer we set $d=256$ and $d_{ff} = 512$ with $H=16$ heads and dropout of $0.1$.
With these parameter setting, the model is constructed with  $560K$ trainable parameters, and wights only $5.5MB$.

\subsection{Training setup}
We train our model using the \ac{SGD} optimizer, which we found to be the best fit for this task.
We use learning rate of $0.0003$, weight decay of $1e^{-5}$, batch size of $128$ and use Spec Augment to prevent the model from over-fitting.
We use \ac{BCE} loss to train our model.
In order to train our model with constant batch size, we construct training samples by randomly selecting 256 frames from sample of the original data.
At inference time, the model operates on the entire sequence. In cases where the input length is too long, it could be split into shorter segments.

\section{EXPERIMENTAL STUDY}
In this section, we describe the experimental setup, including the train and test datasets, the experiments carried out, and
the results.
\subsection{Datasets}
In order to train the proposed method we generated a supervised dataset to simulate various real-life scenarios according to~\eqref{eq:prooblem_time}. The LibriSpeech \cite{panayotov2015librispeech} dataset  was used as a clean speech dataset. This is a relatively clean dataset and hence, a simple energy based \ac{VAD} is used to label the dataset.
Since the speech data in the LibriSpeech dataset is not balanced in terms of speech to non-speech ratio, the data was balanced by adding silence segments to the speech signals in random locations.

To simulate various  \acp{RIR}, the RIRgenerator \cite{habets2006room} was used with randomly chosen acoustic conditions, such as microphone
and speaker positions,room dimensions, and reverberation level according to Table~\ref{table:noisysetup}.

The additive noises were drawn from the WHAM!~\cite{Wichern2019WHAM}  corpus,  which consist of babble noises recorded in different environments such as, restaurants, cafes, bars, and parks. We added the noises to the reverberant speech signal at random \acp{SNR} in the range of $[-3,20]$~dB.

Finally,  $30,000$ training scenarios were generated for the training dataset,  $6,500$ for the validation and  $6,500$ for the test datasets.
Note, that the speakers and the noises were randomly divided prior to the generation of the datasets, to prevent leakage of the same speakers or noise between training, validation and test phases.

\begin{table}[t]
\caption{Noisy reverberant data specification.}
\label{table:noisy_data}
\centering
\resizebox{0.8\columnwidth}{!}{
\begin{tabular}{@{}lll@{}}
\toprule
                        & x        & \emph{U}{[}4,8{]}                                  \\
Room [m]                & y        & \emph{U}{[}4,8{]}                                  \\
                        & z        & \emph{U}{[}2.5,3{]}                                               \\ \midrule
T\_60 [sec]             &          & \emph{U}{[}0.15, 0.6{]}                           \\ \midrule
                        & x        & $\frac{x_{\text{Room}}}{2}$+\emph{U}{[}-0.5,0.5{]} \\
Mic. Pos. [m]           & y        & $\frac{y_{\text{Room}}}{2}$+\emph{U}{[}-0.5,0.5{]} \\
                        & z        & 1.5                                         \\ \midrule

Sources Pos. [$^\circ$] & $\theta$ & \emph{U}{[}0,180{]}                                \\ \midrule
Sources Distance [m]                   &          & 1+\emph{U}{[}-0.5,0.5{]}                         \\ \midrule
SNR [dB]                    &        & \emph{U}{[}-3, 20{]}                                 \\ \bottomrule
\end{tabular}}
\label{table:noisysetup}
\end{table}



\subsection{Experimental setup}
\noindent\textbf{Objective measurements}
We use the \ac{AUC} of the \ac{ROC}, and the \ac{EER} as evaluation metrics for our model.

\noindent\textbf{Compared methods}
We compared our model with 3 other models. The first is the \ac{bDNN} \cite{zhang2015boosting} architecture, which uses an ensemble of \acp{DNN}, each with a different context length, to deal with the problem of a single fixed context length. The second is the  \ac{ACAM} \cite{kim2018voice} and the third is the  \ac{SA} \cite{jo2021self}, which were explained in the introduction.
The parameter count for each of the models is summarized in Table \ref{table_num_param_others}.

\begin{table}[ht]
\caption{Models $\#$ of  parameter}
\begin{tabularx}{0.49\textwidth}{  >{\centering\arraybackslash}X 
   >{\centering\arraybackslash}X 
   >{\centering\arraybackslash}X 
   >{\centering\arraybackslash}X
   >{\centering\arraybackslash}X  } 
 \hline
    & \ac{bDNN} & \ac{ACAM}  & \ac{SA}  & Ours \\
 \hline
  parameters &  1.6M & 870K & 370K & 560K \\ 
 
 \hline

\end{tabularx}
\label{table_num_param_others}
\end{table}

\subsection{Results}
\noindent\textbf{Real-life benchmark}
To test our model, two benchmarks are used. The first, recently published in~\cite{jo2021self}, is constructed with real world MOVIE dataset, containing  audio signals of $6$ movies together with their corresponding \ac{VAD} labels.
The second, is the ENVIRONMENT test set~\cite{kim2018voice}. This test set was recorded at $4$ different environments including:  park,  bus-stop, construction-site and room. 
 
It is worth noting, that for a fair comparison, the reported results of the compared models are the ones which preformed better, on most of the test set in~\cite{jo2021self}. Furthermore, we emphasise that the compared methods are trained and evaluated with $fs=16$~KHz, while our model is trained and evaluated with $fs=8$~KHz by down-sampling the test sets to $8kHz$.

Table~\ref{table_result_movie} shows the \ac{AUC} results of the proposed method together with the 3 compared methods on the MOVIE (6 movies) and ENVIRONMENT (4 environments) test sets. The \ac{SA} algorithm outperforms the \ac{ACAM} and the \ac{bDNN} methods. This states the benefit of this approach. It is further evident that our model, which combines the a CNN embedder before the SA encoder, gains even better results.  Although our model is marginally larger than of the SA model, in inference time, our model is applied on the entire signal at once, while the SA model is applied on each frame separately.


\begin{table}[ht]
\caption{Real world data-AUC(\%)}
\begin{tabularx}{0.49\textwidth}{  >{\raggedright\arraybackslash}X 
   >{\raggedright\arraybackslash}X 
   >{\raggedright\arraybackslash}X 
   >{\raggedright\arraybackslash}X 
   >{\raggedright\arraybackslash}X 
   >{\raggedright\arraybackslash}X 
   >{\raggedright\arraybackslash}X 
   >{\raggedright\arraybackslash}X 
   >{\raggedright\arraybackslash}X  } 
 \toprule
  data\textbackslash model  & \ac{bDNN} & \ac{ACAM}  & \ac{SA} & Ours \\ 
 \toprule
 
 Armageddon & 87.85 & 84.64 & 87.93 & \textbf{88.36}  \\ 
 \hline
 Dead Poets Society & 86.3 & 85.66 & \textbf{86.47} & 83.4  \\ 
 \hline
 Forrest Gump & 87.19 & 86.48 & 87.06 & \textbf{87.39} \\ 
 \hline
 Independence Day & 82.45 & 80.73 & 82.89 & \textbf{83.47} \\ 
 \hline
 Legally Blonde & 81.81 & 81.92 & 81.64 & \textbf{85.93}  \\ 
 \hline
 Saving Private Ryan & 84.65 & 82.38 & 85.23 & \textbf{86.80}  \\ 
 \toprule
  Park & 98.61 & 98.49 & 98.80 & \textbf{99.04}  \\ 
 \hline
  Bus stop & 98.30 & 98.20 & 98.33 & \textbf{98.87}  \\ 
 \hline
  Construction site & 99.66 & 99.48 & 99.68 & \textbf{99.82}  \\ 
 \hline
  Room & 99.06 & 99.02 & 99.20 & \textbf{99.37}  \\ 
 \toprule

\end{tabularx}
\label{table_result_movie}
\end{table}

\noindent\textbf{Ablation study}
The proposed algorithm consists of a \ac{CNN} embedder and a \ac{SA} encoder. In this section compare our model to 2 variants of the proposed architecture, the first is an end-to-end \ac{FCNN} architecture, while the second (Encoder) is an end-to-end \ac{SA} based encoders architecture.  All models were trained with the same training dataset described in the previous section. 
The different variants were designed to have approximately the same number of parameters. Their parameter size are summarized in Table \ref{table_num_param}.

As a test set for this experiment we  used 3 testing sets.
The first is the test set created from the Librispeech data set and the WHAM! noises described in the datasets section.
The second is Disk 6 of the NIST dataset~\cite{doddington2000nist}, commonly used for diarization of phone calls.
The third is a testing set created using the Timit dataset~\cite{garofolo1993timit}, with the WHAM! noise test set, containing multiple speakers in each recording.

\begin{table}[ht]
\caption{Models $\#$ of parameter}
\begin{tabularx}{0.49\textwidth}{  >{\centering\arraybackslash}X 
   >{\centering\arraybackslash}X 
   >{\centering\arraybackslash}X 
   >{\centering\arraybackslash}X  } 
 \hline
    & \ac{FCNN} & Encoder & Ours  \\ 
 \hline
 parameters &  700K & 560K & 560K \\ 
 
 \hline

\end{tabularx}
\label{table_num_param}
\end{table}

All networks were designed to operate on $256$ frames, and receive the same input, the Mel-spectrum.
Although our model is able to operate on the entire sequence at once, and utilize the attention to benefit from the long context, for a fair comparison between the models we show the preference of the proposed model working on $256$ frames at a time.
We also apply the model on the entire input sequence to demonstrate the benefit of long context.

The \ac{EER} and \ac{AUC} results are summarized in Table \ref{table_result}.

\begin{table}[ht]
\caption{EER(\%)/AUC(\%)}
\begin{tabularx}{0.49\textwidth}{  >{\raggedright\arraybackslash}X 
   >{\raggedright\arraybackslash}X 
   >{\raggedright\arraybackslash}X 
   >{\raggedright\arraybackslash}X 
   >{\raggedright\arraybackslash}X  } 
 \toprule
  data\textbackslash model & \ac{FCNN} & Encoder & Ours (256 frames) & Ours entire sequence \\ 
 \toprule
 
 Libri  & \scriptsize 11.68/95.7 & \scriptsize 10.41/96.62 & \scriptsize 5.299/98.94 & \scriptsize \textbf{5.05/99.02} \\ 
 \hline
 Timit & \scriptsize 15.03/92.43 & \scriptsize 14.75/92.22 & \scriptsize 9.723/96.01 & \scriptsize \textbf{8.93/96.54}
 \\ 
 \hline
 Nist  & \scriptsize 10.31/95.17 & \scriptsize 14.5/92.88
 & \scriptsize 9.983/95.12 & \scriptsize \textbf{9.68/95.2}
 \\ 
 \toprule

\end{tabularx}
\label{table_result}
\end{table}

Evidently, the combined architecture outperforms each of its individual components.
Furthermore, using the entire input sequence outperforms the same model using  $256$ frames separately.

\begin{figure}[ht]
    \centering
    \includegraphics[width=0.35\textwidth]{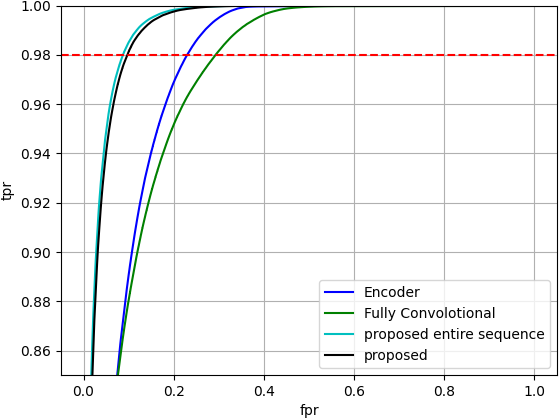}
    \caption{ROC on the Libri test set}
    \label{fig:roc}
\end{figure}

Finally,
the \ac{ROC} of the proposed model with its variants are depicted in \ref{fig:roc}. For a true-positive-rate at  $0.98$, marked with dashed red line,  we show the advantage of the proposed method over its two variants.

\begin{figure}[h]
     \centering
     \begin{subfigure}[b]{0.41\textwidth}
         \centering
         \includegraphics[width=\textwidth]{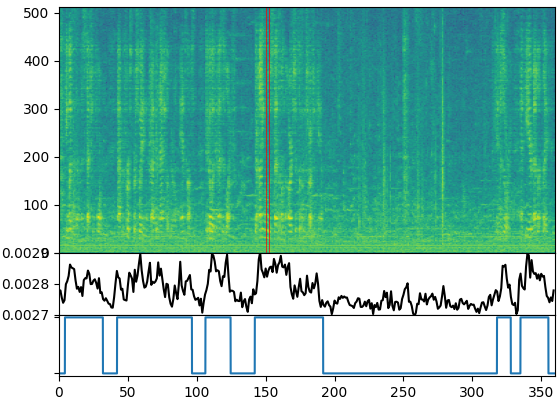}
         \caption{speech}
         \label{fig:attention speech}
     \end{subfigure}\\
     \begin{subfigure}[b]{0.41\textwidth}
         \centering
         \includegraphics[width=\textwidth]{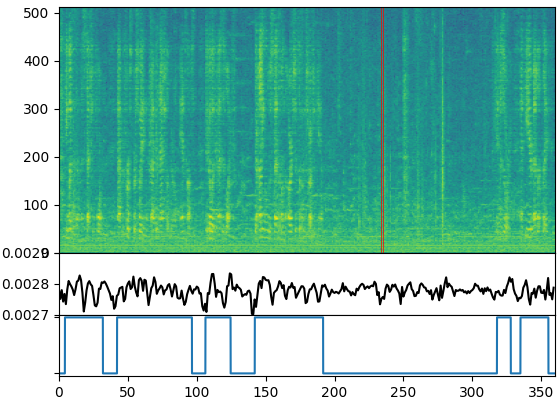}
         \caption{noise}
         \label{fig:attention noise}
     \end{subfigure}\\
     \begin{subfigure}[b]{0.41\textwidth}
         \centering
         \includegraphics[width=\textwidth]{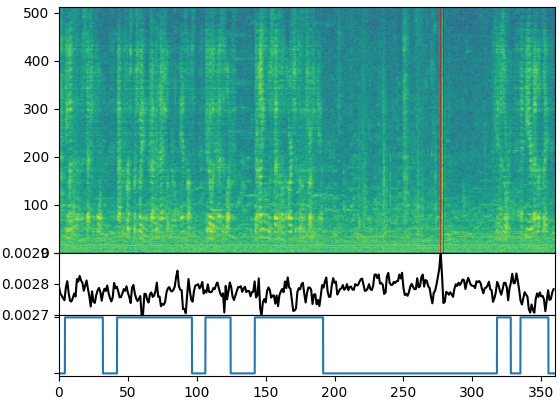}
         \caption{transient noise}
         \label{fig:attention trensient}
     \end{subfigure}
     \hfill
    \caption{Multi-Head Attention of speech (a) noise (b) and transient noise (c) to all other frames.}
    \label{fig:atten}
\end{figure}

\noindent\textbf{Attention visualization}
To further evaluate the attention operation of the proposed architecture,  a test sample from the LibriSpeech test-set is used.  We visualize the average multi-head attention wights~\eqref{eq:AverageAttention} with respect to a speech frame, a noise frame and a transient noise frame. Fig. \ref{fig:attention speech} depicts~\eqref{eq:AverageAttention} for a speech frame. Fig. \ref{fig:attention noise} for a noise frame, and Fig. \ref{fig:attention trensient} for a transient noise frame.
In each Figure, the frame of interest is signed with a red rectangle on the Spectrogram, the black line plot represents \eqref{eq:AverageAttention} and the blue line plot represents the \ac{VAD} true lable.

Interestingly, for the speech frame, the encoder paid attention to frames in which speech was active, and vice-versa to noise only frames. It is evident that even 'far' speech is beneficial for the current speech frame.  


For the noise frame, it is easy to see  that since the noise is non-stationary and uncorrelated over time, the attention to all frames was low.

Finally, for the transient noise, it is evident that the highest attention was set to the same frame, since it is different than most other frames, noise and speech.

\section{Conclusion}
A CNN-SA architecture was presented for detecting speech given a noisy reverberant signal. CNN embedder is first used to exploit the spatial information of the input spectrum. The \ac{SA} encoder is then applied on the frame-wise embedding sequence to gain contextual information. Finally, the model outputs the estimation of the presence of the speech. Experiments show that the proposed architecture outperforms end-to-end \ac{FCNN} as well as \ac{SA} architectures. The proposed method also achieve \ac{SOTA} results on two real-life benchmarks.


 \balance
\bibliographystyle{IEEEtran}
\bibliography{ref}

\end{document}